\documentclass[pra,aps,twocolumn,superscriptaddress,showpacs,10pt]{revtex4-1}

\usepackage{graphicx}
\usepackage{amsmath,amssymb,times}
\usepackage{subfigure}
\usepackage{bbm}
\usepackage{color}

\newcommand{\ket}[1]{|#1\rangle}

\begin{document}

\title{Hybrid methods for witnessing entanglement in a microscopic-macroscopic system}

\author{Nicol\`{o} Spagnolo}
\affiliation{Dipartimento di Fisica, Sapienza Universit\`{a} di Roma, piazzale Aldo Moro 5, 00185 Roma, Italy}
\affiliation{Consorzio Nazionale Interuniversitario per le Scienze Fisiche della Materia, piazzale Aldo Moro 5, 00185 Roma, Italy}
\author{Chiara Vitelli}
\affiliation{Dipartimento di Fisica, Sapienza Universit\`{a} di Roma, piazzale Aldo Moro 5, 00185 Roma, Italy}
\author{Mauro Paternostro}
\affiliation{School of Mathematics and Physics, Queen's University, Belfast BT7 1NN, United Kingdom}
\author{Francesco De Martini}
\affiliation{Dipartimento di Fisica, Sapienza Universit\`{a} di Roma, piazzale Aldo Moro 5, 00185 Roma, Italy}
\affiliation{Accademia Nazionale dei Lincei, via della Lungara 10, I-00165 Roma, Italy}
\author{Fabio Sciarrino}
\email{fabio.sciarrino@uniroma1.it}
\affiliation{Dipartimento di Fisica, Sapienza Universit\`{a} di Roma, piazzale Aldo Moro 5, 00185 Roma, Italy}
\affiliation{Istituto Nazionale di Ottica, Consiglio Nazionale delle Ricerche (INO-CNR), largo E. Fermi 6, I-50125 Firenze, Italy}

\begin{abstract}
We propose a hybrid approach to the experimental assessment of the genuine quantum features of a general system consisting of microscopic and macroscopic parts. We infer entanglement by combining dichotomic measurements on a bidimensional system and phase-space inference through the Wigner distribution associated with the macroscopic component of the state. As a benchmark, we investigate the feasibility of our proposal in a bipartite-entangled state composed of a single-photon and a multiphoton field. Our analysis shows that, under ideal conditions, maximal violation of a Clauser-Horne-Shimony-Holt-based inequality is achievable regardless of the number of photons in the macroscopic part of the state. The difficulty in observing entanglement when losses and detection inefficiency are included can be overcome by using a hybrid entanglement witness that allows efficient correction for losses in the few-photon regime.
\end{abstract}


\maketitle

\section{Introduction}

An open challenge for fundamental quantum physics is to affirm the quantum nature of a system that puts together a microscopic part and a mesoscopic one. This hybrid scenario can emerge in completely different experimental platforms ranging from individual spin systems interacting with multimode cavity fields, such as transmon qubits in coplanar transmission-line resonators~\cite{Wall04,Aoki06}, to ionic impurities embedded in ultracold atomic samples, such as the systems considered in some recent experiments reported in~\cite{Zipk10,Schm10}. Another possible physical approach exploits a massive tiny mirror interacting optomechanically with a single photon within a Michelson interferometer \cite{Giga06,Arci06,Klec06,DeMa10,Mars03}. This endeavor could contribute to challenge the observability of quantum features at the macroscopic level, which is one of the most fascinating open problems in quantum physics. The difficulties inherent in such a quest are manifold, and they are related on the one hand to the unavoidable interaction of the system with the surrounding environment \cite{Zure03,Kwia01,Pan03,Pete04}. On the other hand, one faces the debated problem of achieving a measurement precision sufficient to observe quantum effects at such macroscales \cite{Kofl08,JPR08}. In this context, it has been experimentally proven that a dichotomic measurement performed upon a multiphoton-entangled state is not sufficient to catch quantumness \cite{Vite10}. The accuracy of the measurement is crucial for the observation of quantum features and should be put on the same footing as the use of proper entanglement and nonlocality criteria for macroscopic quantum systems \cite{JPR08,Wodk00,Gied01,Stob07,Lee09,Lee10}.

To successfully tackle the manipulation and characterization of hybrid systems the following question is still open: How can we ascertain the nonclassical nature of a multipartite state that, per se, does not meet the criteria for quantumness that have been designed for system components of equal dimensionality? Our work provides a quantitative answer to this broad question. We introduce an investigative platform that can be built up without the necessity for information on the state itself, and this supports the general validity and broad applicability of our results.

We introduce a hybrid method to demonstrate experimentally the truly quantum mechanical features of a general microscopic-macroscopic system beyond any assumption on its state and without the necessity of any a priori state knowledge. We infer the entanglement properties by means of a hybrid approach that combines dichotomic measurements on a bidimensional system and phase-space inferences through the Wigner distribution associated with the macroscopic component of the state. Here, through the use of a hybrid entanglement test, we identify a valuable tool for our goals. While the microscopic part of the state is measured using spin-$1/2$ projection operators, the macroscopic counterpart undergoes phase-space measurements based on the properties of itsWigner function~\cite{Wodk00}. At variance with previous proposals \cite{Bana99,Wodk00}, the approach presented in this paper is tailored to fully exploit the polarization-spin degree of freedom on both the microscopic and the macroscopic subsystems. We analyze the effects of losses on a Clauser-Horne-Shimony-Holt-like (CHSH-like) inequality test~\cite{Clau69} and show that maximum violation is achieved when losses are absent, regardless of the size of the macroscopic part of the state. This is not the case under nonideal conditions. However, we show how losses can be efficiently taken into account so as to infer entanglement of our multiphoton state.

As a paradigmatic microscopic-macroscopic system (MMS), we investigate the state obtained from a fully microscopic-entangled system through an amplification process~\cite{Dema08,Spag10}.  Such system has been further considered recently as a benchmark to perform nonlocality tests with human-eye threshold detectors \cite{Seka09} or as a platform for absolute radiometry \cite{Sang10}. At variance with respect to Refs.~\cite{Dema08,Spag10}, our approach does not require any assumption on the system under investigation and hence represents a genuine entanglement test.

The present paper is organized as follows. In Sec. \ref{sec:hybrid_nonloc} we introduce and define the CHSH-based entanglement inequality based on hybrid measurements for the single-photon and the multiphoton modes. Then, in Sec. \ref{sec:hybrid_ent} we discuss how the CHSH-based test defined in Sec. \ref{sec:hybrid_nonloc} can be modified to obtain an entanglement witness tailored to be applied in a lossy scenario. Finally, in Sec. \ref{sec:hybrid_benchmark} we provide a specific example of a joint optical system composed by a single-photon and a multiphoton field based on the process of optical parametric amplification. We then run both the CHSH-based test and the entanglement witness on this system in order to identify in which range of the system's parameters the entanglement can be addressed with our approach.

\section{Hybrid entanglement test based on Bell's inequalities}

\label{sec:hybrid_nonloc}

\begin{figure}[ht!]
\centering
\includegraphics[width=0.45\textwidth]{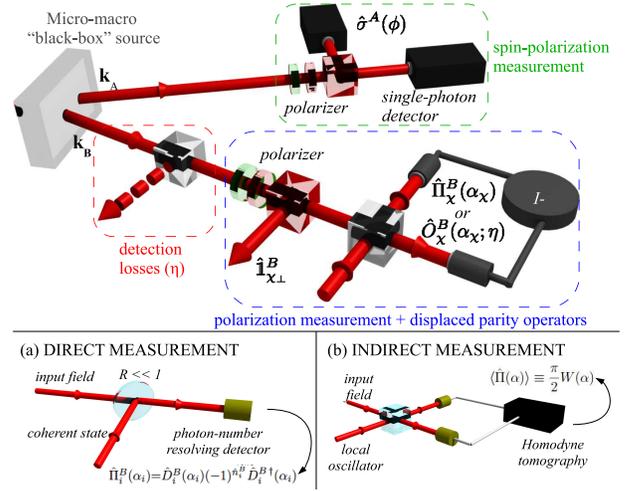}
\caption{(Color online) Hybrid entanglement test on an optical MMS state generated by a ``black-box.'' The single-photon mode $\mathbf{k}_{A}$ is measured by a polarization-detection apparatus, while the multiphoton mode $\mathbf{k}_{B}$ undergoes polarization projection and the measurement of the displaced parity operators. (a) Direct measurement of the $\hat{\Pi}(\alpha)$ displaced parity operators. (b) Indirect measurement of the average value $\langle \hat{\Pi}(\alpha) \rangle$ of the displaced parity operators by exploiting a homodyne-detection apparatus.}
\label{fig:schemaConceptual}
\end{figure}

Let us consider a general MMS state with its microscopic part embodied by a single-photon polarization state (a qubit). We take the macroscopic part, on the other hand, as encoded in the multiphoton state of a continuous-variable (CV) system. The two subsystems are supposed to be entangled by a mechanism whose details are inessential for our tasks here. A benchmark state of such situation will be provided later. Polarization measurements performed over the state of the single-photon mode $\mathbf{k}_{A}$ are described by the Pauli spin operator $\hat{\sigma}^{A}(\phi)= \vert \phi \rangle_{A} \langle \phi \vert - \vert \phi_{\bot} \rangle_{A} \langle \phi_{\bot} \vert$, where $\phi$ is the direction identifying the polarization state in the Poincar\'e sphere and $\phi_\bot$ is its orthogonal direction. The CV measurements, on the other hand, are given by $\hat{\Pi}_{\chi,\chi_{\bot}}^{B}(\alpha_{\chi},\chi){=}\hat{\Pi}^{B}_{\chi}(\alpha_{\chi}) {\otimes}\hat{\mathbbm{1}}^{B}_{\chi_{\bot}}$, where $\hat{\Pi}_{i}^{B}(\alpha_{i}){=}\hat{D}_{i}^{B}(\alpha_{i}) (-1)^{\hat{n}_{i}^{B}} \hat{D}_{i}^{B\, \dag}(\alpha_{i})$ is the displaced parity operator built from the displacement $\hat{D}^B_i(\alpha_i)$ ($\alpha_i{\in}\mathbb{C}$) and the number operator $\hat{n}^{B}_{i}$ ($i{=}\{\chi,\chi_{\bot}\}$ stands for the polarization state. Such operators can be directly measured \cite{Dele08,Laih10} by combining the input field with a coherent state in a low reflectivity beam splitter and by measuring the parity of the output field  [see Fig. \ref{fig:schemaConceptual} (a)]. However, such technique requires a photon-counting technique with very high efficiency, a condition extremely difficult to achieve with the present technology.

%% FIGURE 1 %%
%

An indirect measurement of the average value of the displaced parity operators can be performed by exploiting the connection between $\langle \hat{\Pi}(\alpha) \rangle$ and the Wigner function of the state [see Fig. \ref{fig:schemaConceptual} (b)]. Indeed, the average value of the measurement operator on state $\rho^B_i$ of the multiphoton mode is related to the value of its Wigner function at $\alpha_i$, $W^{B}_{\Phi}(\alpha_{i}){=}(2/\pi)\text{Tr}[\hat{\Pi}_{i}^{B}(\alpha_{i})\rho^B_{i}]$. The latter can be easily reconstructed using homodyne measurements. 
We define the qubit-CV correlator ${\cal C}(\alpha_{\chi},\chi; \phi){=} \langle\hat{\sigma}^{A}(\phi){\otimes}\hat{\Pi}_{\chi,\chi_{\bot}}^{B}(\alpha_{\chi},\chi)\rangle$, which is evaluated on a general MMS $\rho_{AB}$, and the CHSH-based entanglement parameter
\begin{equation}
\label{eq:CHSH_def}
%\begin{aligned}
\mathcal{B} {=} \mathcal{C}(\alpha'_{\chi},\chi';\phi'){+}\mathcal{C}(\alpha'_{\chi},\chi';\phi)
{+}\mathcal{C}(\alpha_{\chi},\chi;\phi'){-}\mathcal{C}(\alpha_{\chi},\chi;\phi).
%\end{aligned}
\end{equation}
A more detail discussion can be found in Appendix \ref{sec:appA}. As the average values of the outcomes of the $\hat{\sigma}^{A}(\phi)$ and $\hat{\Pi}_{\chi,\chi_{\bot}}^{B}(\alpha_{\chi},\chi)$ measurements is limited by $\langle \hat{\sigma}^{A}(\phi) \rangle \leq 1$ e $\langle \hat{\Pi}_{\chi,\chi_{\bot}}^{B}(\alpha_{\chi},\chi) \rangle \leq 1$, for all separable states the bound $\vert \mathcal{B}_{} \vert{ \leq}2$ holds. A violation of this bound witnesses an entangled state. The measurement settings for the single-photon mode $\mathbf{k}_{A}$ [multiphoton mode $\mathbf{k}_{B}$] are given by the measured polarizations ($\phi$, $\phi'$) [measured polarizations ($\chi$, $\chi'$) and the chosen phase-space points ($\alpha_{\chi}$, $\alpha'_{\chi}$)]. This requires a standard polarization detection system for the microscopic mode and a homodyne detection system for the multiphoton one, as shown in the scheme presented in Fig.~\ref{fig:schemaConceptual} (b).

We conclude by observing that the inequality of Eq.(\ref{eq:CHSH_def}) becomes a nonlocality test when the displaced parity operator are directly measured on the multiphoton field [see Fig. \ref{fig:schemaConceptual} (a)], since no assumption are necessary on the detection apparatus. In this case, the outcome of the $\hat{\sigma}^{A}(\phi)$ and $\hat{\Pi}_{\chi,\chi_{\bot}}^{B}(\alpha_{\chi},\chi)$ measurements can only be $\pm{1}$, and the use of a local-hidden-variable (LHV) model imposes the bound $\vert \mathcal{B}_{} \vert{ \leq}2$ \cite{Clau69} on the $\mathcal{B}$ parameter. A violation of this bound confutes all LHV theories. 

\section{Hybrid entanglement witness with losses}

\label{sec:hybrid_ent}

The test presented above can be modified so as to embody a witness able to reveal entanglement when the state at hand is affected by losses. This is modeled by inserting a beam splitter of transmittivity $\eta{\in}[0,1]$ in the path of the modes at hand, ``tapping" the corresponding signal~\cite{JPR08}. The choice $\eta{=}1$ ($\eta{=}0$) corresponds to a lossless (fully lossy) process. To this end, the measurement performed on the $\vec{\pi}_{\chi}$ polarization of the multiphoton part is replaced by the operator~\cite{Lee10}
\begin{equation}
\label{eq:operator_witness}
\hat{O}_{\chi}^{B}(\alpha_{\chi};\eta){=}\left\{\!\begin{array}{ll} \frac{1}{\eta} \hat{\Pi}_{\chi}^{B}(\alpha_{\chi}){+}\left( 1{-} \frac{1}{\eta} \right){\hat{\mathbbm{1}}_{\chi}^{B}}& \mathrm{if} \ \eta{\in}(0.5,1], \\ 2 \hat{\Pi}_{\chi}^{B}(\alpha_{\chi}){-}\hat{\mathbbm{1}}_{\chi}^{B} & \mathrm{if} \ \eta{\in}(0,0.5].\end{array} \right.
\end{equation}
In this way, the overall measurement on the macroscopic subsystem reads $\hat{O}_{\chi,\chi_{\bot}}^{B}(\alpha_{\chi},\chi;\eta){=}\hat{O}^{B}_{\chi}(\alpha_{\chi};\eta){\otimes}\hat{\mathbbm{1}}^{B}_{\chi_{\bot}}$. For any separable state being measured after the lossy process, $\vert \langle \hat{O}_{\chi,\chi_{\bot}}^{B}(\alpha_{\chi},\chi;\eta) \rangle_{\eta} \vert\leq1$~\cite{Cahi69a,Cahi69b,Lee09,Lee10}. Hence, by introducing the MMS correlator ${\tilde{\cal C}}_\eta(\alpha_{\chi}, \chi,\phi){=}\langle \hat{\sigma}^{A}(\phi){\otimes}\hat{O}_{\chi,\chi_{\bot}}^{B}(\alpha_{\chi},\chi;\eta) \rangle_{\eta}$, we define 
\begin{equation}
\label{eq:witness_def}
\begin{aligned}
\mathcal{W}_{\eta}&{=}\tilde{\mathcal{C}}_{\eta}(\alpha'_{\chi},\chi',\phi'){+}\tilde{\mathcal{C}}_{\eta}(\alpha'_{\chi},\chi',\phi){+}\\
&{+}\tilde{\mathcal{C}}_{\eta}(\alpha_{\chi},\chi,\phi'){-}\tilde{\mathcal{C}}_{\eta}(\alpha_{\chi},\chi,\phi).
\end{aligned}
\end{equation}
Any separable state undergoing a lossy process on mode $\mathbf{k}_{B}$ is bound to satisfy $\vert \mathcal{W}_{\eta}^{sep} \vert{\leq}2$ (see Appendix \ref{sec:appB}). Violation of this inequality witnesses entanglement in the system. Such a bound can be explained by considering that separable states do not violate CHSH inequalities, and local processes such as losses cannot increase their nonlocal character. It is important to notice that, by virtue of the assumption that the macrostate of mode $\mathbf{k}_{B}$ undergoes losses $\eta$ before (rather than {\it at}) detection, this entanglement witness reveals the presence of entanglement \textit{without} any assumption on the MMS source (see Fig.\ref{fig:schemaConceptual}). On the other hand, the lossy mechanism can be shifted to occur just before measurement, thus modeling the effects of a non-ideal detector. For $\eta{=}1$, $\mathcal{W}_{\eta}$ coincides with the CHSH-based parameter $\mathcal{B}$ in Eq.~(\ref{eq:CHSH_def}).

%% FIGURE 2 %%
%
\begin{figure}[b!]
\centering
\includegraphics[width=0.45\textwidth]{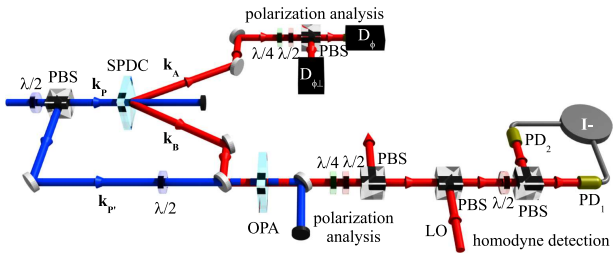}
\caption{(Color online) Layout of the MMS source based on the process of optical parametric amplification of a single photon belonging to an entangled pair.}
\label{fig:schemaNonlocality}
\end{figure}

\section{Experimental benchmark}

\label{sec:hybrid_benchmark}

In this section, we analyze in details a specific optical system to evaluate the effectiveness of our hybrid approach. As a benchmark for the hybrid CHSH-based entanglement test and entanglement witness described above, we analyze the MMS-state-source addressed in Ref.~\cite{Dema08}. A layout of the system is reported in Fig.~\ref{fig:schemaNonlocality}. 
The polarization singlet state $\vert \psi^{-} \rangle_{AB}{=}(\vert H \rangle_{A} \vert V \rangle_{B}{-}\vert V \rangle_{A} \vert H \rangle_{B})/\sqrt 2$ of a photon pair is generated in a nonlinear crystal through a spontaneous parametric down-conversion (SPDC) process. Here $\ket{H}$ ($\ket{V}$) stands for the horizontal (vertical) polarization state. The photon populating mode $\mathbf{k}_{B}$ is then injected into an optical parametric amplifier (OPA) in a collinear configuration. Since the OPA implements a unitary operation, the symmetry of $\vert \psi^{-} \rangle_{AB}$ is preserved by the amplification process and the overall state $\vert \Psi^{-} \rangle_{AB}{=}(\vert \phi \rangle_{A} \vert \Phi^{\phi}_{\bot} \rangle_{B}{-}\vert \phi_{\bot} \rangle_{A} \vert \Phi^{\phi} \rangle_{B})/\sqrt{2}$ maintains rotational invariance form for any polarization basis. Here, $\vert \Phi^{\phi} \rangle$ are the multiphoton states generated by amplification of a single-photon polarization state $\vert \phi \rangle$. Quantum entanglement between the micropart and the macropart of $\vert \Psi^{-} \rangle_{AB}$  has been demonstrated~\cite{Dema08} under a supplementary assumption on the source~\cite{Spag10}. The OPA performs the optimal cloning process only for equatorial polarization $\vec{\pi}_{\phi}{=}(\vec{\pi}_{H}{+}e^{\imath \phi} \vec{\pi}_{V})/\sqrt 2$. We thus restrict our attention to this subset of polarization states, which motivates our choice for $\hat{\sigma}^A(\phi)$ performed above. 

%% FIGURE 3 %%
%
\begin{figure}[ht!]
\centering
\includegraphics[width=0.35\textwidth]{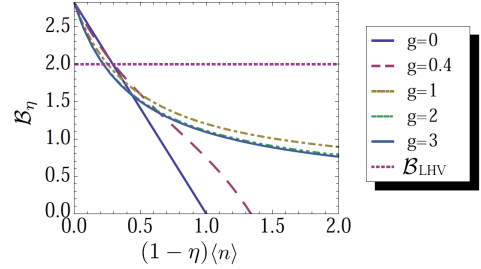}
\caption{(Color online) CHSH-based parameter $\mathcal{B}_{\eta}$ as a function of the number of lost photons $(1-\eta) \langle n \rangle$ for different values of the gain $g$. We show the local realistic boundary ${\cal B}_\text{LHV}{=}2$.}
\label{fig:CHSH_B}
\end{figure}
%
%% FIGURE 4 %%
%
\begin{figure}[b!]
\centering
\includegraphics[width=0.45\textwidth]{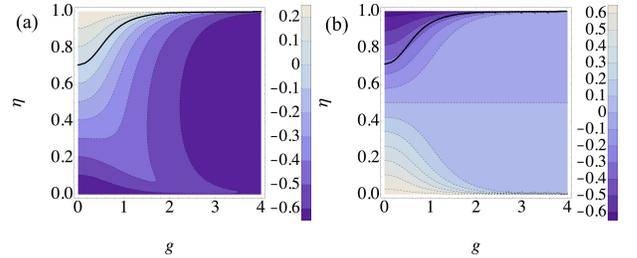}
\caption{(Color online) (a) Contour plot of the shifted loss function $\mathcal{L}(g,\eta){-}{2}^{-1/2}$ as a function of the gain $g$ and the detection efficiency $\eta$. (b) Contour plot of the negativity of the Wigner function of an amplified single-photon state~\cite{Spag09} against $g$ and $\eta$, evaluated at the origin of the phase space. In both panels the solid line divides the region of entanglement ($\vert \mathcal{B}_{\eta} \vert{>}2$, above the line) from the one in which entanglement cannot be inferred ($\vert \mathcal{B}_{\eta}\vert \leq 2$, below the line).}
\label{fig:CHSH_B_3D}
\end{figure}
%
%% FIGURE 5 %%
%
\begin{figure*}[ht!]
\centering
\includegraphics[width=0.95\textwidth]{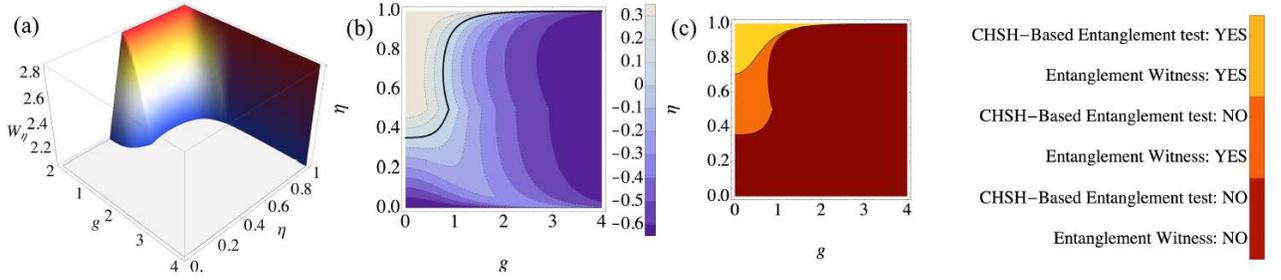}
\caption{(Color online) (a) Plot of $W_{\eta}$ vs	 the detection efficiency $\eta$ and the nonlinear gain $g$. (b) Contour plot of the effective loss function $h(\eta) \mathcal{L}(\eta,g)$. Entanglement can be revealed in the region above the black line. (c) Summary of the results obtained from our tests. We identify three regions in the $(\eta,g)$ space, depending on whether entanglement can be demonstrated with our techniques.}
\label{fig:W_3D}
\end{figure*}

We now discuss the results of the CHSH-based test and the application of the entanglement witness to the MMS state given in Fig.~\ref{fig:schemaNonlocality}. We begin analyzing the CHSH-based inequality (\ref{eq:CHSH_def}) in the lossless case ($\eta{=}1$). The correlation operator evaluated on $\vert \Psi^{-} \rangle_{AB}$ takes the form~(see Appendix \ref{sec:appC})
\begin{equation}
\label{eq:correlator_lossless}
\mathcal{C}(X_{\chi},P_{\chi},\chi;\phi){=}(1{-}{\cal Z})\cos[2(\chi{-}\phi)] e^{-{\cal Z}},
\end{equation}
where ${\cal Z}{=}2(e^{-2 g} \overline{X}^{2}_{\chi}{+}e^{2 g} \overline{P}^{2}_{\chi})$ is a function of the rotated variables $\overline{X}_{\chi}{=}X_{\chi} \cos(\chi/2){-}P_{\chi} \sin(\chi/2)$ and $\overline{P}_{\chi}{=}X_{\chi} \sin(\chi/2){+}P_{\chi} \cos(\chi/2)$. $(X_{\chi},P_{\chi})$ are the field quadratures and $\alpha_{\chi}{=}X_{\chi}{+}\imath P_{\chi}$. The correlator in Eq.~(\ref{eq:correlator_lossless}) is maximized at the origin of the phase space, where $\mathcal{C}(0,0,\chi;\phi){=}\cos[2(\chi - \phi)]$, which is independent of the gain of the amplifier $g$ and the number of generated photons $\overline{n}=\sinh^2 g$. The correlator has the same form as a Bell-CHSH test performed on a polarization photon pair, where spin-$1/2$ operators are measured. The CHSH-based parameter $\mathcal{B}$ is then maximized by choosing the measurement settings for $(\phi,\phi,\chi,\chi')$ corresponding to such case, which ensures the maximum degree of violation of the local realistic boundary, i.e., ${\cal B}{=}2\sqrt 2$.

We are now in a position to address the possibility to observe MMS entanglement under realistic experimental conditions. We thus analyze the effects of detection efficiency at the homodyne apparatus, while other sources of experimental imperfections, as well as a more detailed derivation, are discussed in Appendix \ref{sec:appD}). The measurement of the generalized parity operator on the multiphoton mode $\mathbf{k}_{B}$ can be performed using homodyne detection. Furthermore we include in the qubit-CV correlator the possibility of a nonunitary detection efficiency on mode ${\bf k}_B$ (see Appendix \ref{sec:appD}). By restricting our attention to the origin of the phase space, where maximum nonclassical effects are achieved, we get
$\mathcal{C}^{}_{\eta}(0,0,\chi; \phi){=}\cos[2(\chi{-}\phi)] \mathcal{L}(g,\eta)$, where 
\begin{equation}
\mathcal{L}(g,\eta){=}\frac{\eta[1+2\overline{n}(1-\eta)]}{(1+4 \eta (1-\eta) \overline{n})^{3/2}} 
\end{equation}
is a loss function for the test. Hence, the maximum amount of violation is directly determined by the loss function as $\mathcal{B}_{\eta}{=}\mathcal{B} \mathcal{L}(g,\eta)$.
In Fig.~\ref{fig:CHSH_B} we show the value of $\mathcal{B}_{\eta}$ as a function of the average number of lost photons, $(1{-}\eta) \langle{n}\rangle$, where $\langle{n}\rangle=3\overline{n}+1$ is the mean number of the generated photons after the amplification process. The CHSH-based inequality of Eq.~(\ref{eq:CHSH_def}) is satisfied when only a moderate number of photons is lost. A lower bound $\eta_{lim}{=}1/\sqrt 2$ for the detection efficiency can be found below, where a violation is no longer observed. On the other hand, at set values of $\eta$ there is a minimum gain $g_{lim}(\eta)$ above which the presented test cannot detect MMS-entangled correlations. Such threshold value decreases with the reduction of the efficiency $\eta$. The behavior of $\mathcal{B}_{\eta}$ in the $(\eta,g)$-plane is shown by the contour plot in Fig.~\ref{fig:CHSH_B_3D} (a). In order to relate the violation of the CHSH-based inequality to intrinsically nonclassical features enforced at the level of the macro-part of the state, Fig.~\ref{fig:CHSH_B_3D} (b) reports the negativity of the Wigner function of an amplified single-photon state versus $\eta$ and $g$~\cite{Spag09}. We observe that the transition of $\mathcal{B}_{\eta}$ to the region below the classical limit is directly linked to the decrease in the negativity of the Wigner function itself. Indeed the value of the MMS correlator $\mathcal{C}^{}_{\eta}$  is determined by the excursion of the Wigner function in $X_{\chi}{=}P_{\chi}{=}0$, as a function of the polarization of the injected photon.

We complement the analysis of our MMS by discussing the use of the entanglement witness described above. The evaluation of the correlation operator over state $\vert \Psi^{-} \rangle_{AB}$ after losses leads to $\tilde{\mathcal{C}}_{\eta}^{}(\alpha_{\chi}, \chi,\phi){=}h(\eta) \mathcal{C}_{\eta}^{}(\alpha_{\chi}, \chi; \phi)$, where $h(\eta){=}1/{\eta}$ ($h(\eta){=}2$) for ${1}/{2}{<}\eta{\leq}1$ ($0{\le}\eta{\leq}{1}/{2}$). More details can be found in Appendix \ref{sec:appE}. Therefore, the entanglement witness can be directly obtained from the CHSH-based parameter as $\mathcal{W}_{\eta}{=}h(\eta) \mathcal{B}_{\eta}$. In Fig.~\ref{fig:W_3D} (a) we report the dependence of $\mathcal{W}_{\eta}$ as a function of $\eta$ and $g$. For single-photon states (i.e., at $g=0$), the correction of losses introduced by the factor $h(\eta)$ allows one to observe MMS entanglement up to $\eta{\sim}0.35$. As the number of photons in the macrostate increases, the damping in the negativity of the Wigner function induced by losses scales more rapidly than $\eta$ and the $h(\eta)$-correcting term becomes less effective. Fig.~\ref{fig:W_3D} (b) shows the behavior of the effective overall loss function $h(\eta) \mathcal{L}(\eta,g)$, highlighting the thresholds in $g$ and $\eta$, above which entanglement is observed. We note the non-monotonic behaviour obtained for the inefficiency parameter at $\eta=0.5$, which is a property of the witness itself. However, being Eq. (\ref{eq:operator_witness}) a witness for entanglement, no special meaning can be attached to the lack of violation of the separability condition $\vert \mathcal{W}_{\eta}^{sep} \vert \leq 2$.

\section{Conclusions and perspectives}

We have proposed an experimentally oriented approach to detect entanglement in a MMS-entangled state involving a single-photon and a multiphoton bipartite system. We have used a hybrid CHSH-based inequality and an entanglement witness, whose use against such a class of states is effective. Furthermore, the CHSH-based inequality can be adopted as a genuine nonlocality test when a direct measurement of the displaced parity operators is performed on the multiphoton field. As an experimental benchmark, we applied the proposed inequalities to the bipartite state obtained by amplification of an entangled single-photon-singlet state. While our study spurs further interest in the identification of suitable tests in the high-loss and large-photon-number region, it paves the way to an experimentally feasible demonstration of entanglement properties in an interesting class of states lying at the very border between quantum and classical domains.

\begin{acknowledgments}
We acknowledge support by the FIRB Futuro in Ricerca HYTEQ and Progetto d'Ateneo of Sapienza Universit\`{a} di Roma. M.P. is grateful to the Dipartimento di Fisica, Sapienza Universit\`a di Roma, for hospitality and acknowledges support from EPSRC (EP/G004579/1).
\end{acknowledgments}

\appendix

\section{Hybrid polarization-continuous variables CHSH-based test}

\label{sec:appA}

In this section we review the CHSH-based inequality performed in the paper. Our test is the extension of the Bell's inequality test proposed by Wodkiewicz in Ref. \cite{Wodk00}. We begin by focusing our attention on the multiphoton mode $\mathbf{k}_{B}$. Our MMS, which is generated by amplification of an entangled polarization photon pair, is strongly correlated in such a degree of freedom. To exploit it, we define the measurement operator of the multiphoton state as
\begin{equation}
\hat{\Pi}_{\chi,\chi_{\bot}}^{B}(\alpha_{\chi},\chi) = \hat{\Pi}^{B}_{\chi}(\alpha_{\chi}) \otimes \hat{\mathbbm{1}}^{B}_{\chi_{\bot}}
\end{equation}
Here $\hat{\Pi}^{B}_{i}(\alpha_{i}) = \hat{D}^{B}_{i}(\alpha_{i}) (-1)^{\hat{n}^{B}_{i}} \hat{D}^{B \dag}_{i}(\alpha_{i})$ is the generalized parity operator, where $\hat{D}^{B}(\alpha_{i})$ is the displacement operator and the subscript $i = \{ \chi,\chi_{\bot} \}$ describes the polarization mode. This definition of the measurement operator corresponds to the application of a displacement operator $\hat{D}_{i}^{B}(\alpha_{i})$ followed by a parity measurement. 

In order to detect the correlations present in the system in the polarization degree of freedom, we perform a measurement of the Pauli operator $\hat{\sigma}^{A}(\phi)$ on a single-photon mode. Here, $\hat{\sigma}(\phi)$ is the Pauli $\hat{\sigma}_{z}$ operator along the direction of the Bloch sphere identified by the equatorial polarization state $\vec{\pi}_{\phi} = 2^{-1/2} (\vec{\pi}_{H} + e^{\imath \phi} \vec{\pi}_{V})$. The correlation of the joint system is then defined as
\begin{equation}
\label{eq:corr_definition}
{\cal C}^{}(\alpha_{\chi}, \chi; \phi) = \langle\hat{\sigma}^{A}(\phi) \otimes \hat{\Pi}^{B}(\alpha_{\chi},\chi)\rangle,
\end{equation}
where the averages are evaluated on the investigated $\vert \Psi \rangle_{AB}$ MMS state. Since this correlation operator corresponds to a set of dichotomic measurements, we can use the CHSH-based inequality~\cite{Clau69}
\begin{equation}
\label{eq:CHSH_def_2}
\begin{aligned}
\mathcal{B}{=} \mathcal{C}^{}(\alpha'_{\chi},\chi';\phi'){+}\mathcal{C}^{}(\alpha'_{\chi},\chi';\phi){+ }\mathcal{C}^{}(\alpha_{\chi},\chi;\phi'){-}\mathcal{C}^{}(\alpha_{\chi},\chi;\phi),
\end{aligned}
\end{equation} 
 Here, the measurement settings for the single-photon mode $\mathbf{k}_{A}$ are given by the measured polarizations ($\phi$, $\phi'$), while the measurement settings for the multiphoton mode $\mathbf{k}_{B}$ are given by the measured polarizations ($\chi$, $\chi'$) and the chosen phase space points ($\alpha_{\chi}$, $\alpha'_{\chi}$).

\section{Hybrid polarization-continuous variables entanglement witness with inefficient detectors}

\label{sec:appB}

In this section we discuss in details the hybrid entanglement witness defined in the paper. Such an inequality is an extension of the CHSH-based test of Eq.~(\ref{eq:CHSH_def_2}) where different measurement operators are exploited in the multiphoton mode. The main idea of this extension is to take into account detection losses in order to build measurement operators apt for witnessing entanglement with an inefficient detection apparatus. To this end, the measurement performed on the $\vec{\pi}_{\chi}$ polarization of the multiphoton field can be replaced by the operator \cite{Lee09,Lee10}
\begin{equation}
\hat{O}_{\chi}^{B}(\alpha_{\chi};\eta) = \left\{ \begin{array}{ll} \frac{1}{\eta} \hat{\Pi}_{\chi}^{B}(\alpha_{\chi}) + \left( 1 - \frac{1}{\eta} \right) \hat{\mathbbm{1}}_{\chi}^{B} & \mathrm{if} \ \frac{1}{2} < \eta \leq 1, \\ 2 \hat{\Pi}_{\chi}^{B}(\alpha_{\chi}) - \hat{\mathbbm{1}}_{\chi}^{B} & \mathrm{if} \ \eta \leq \frac{1}{2}, \end{array} \right.
\end{equation}
where $\eta$ is the detection efficiency of the apparatus. Such definition of the measurement operator is performed in order to correct the detrimental effect of losses on the properties of the detected state. Let us consider a general state $\vert \Phi \rangle_{\chi}^{B}$ on spatial mode $\mathbf{k}_{B}$ and polarization $\vec{\pi}_{\chi}$. (Although we illustrate our argument using pure states of mode $B$, our arguments apply equally to mixed states). After losses occur, the state evolves into a density matrix $\hat{\rho}_{\Phi \, \chi}^{\eta \, B}$. The average value of $\hat{O}_{\chi}^{B}(\alpha_{\chi};\eta)$ on such a density matrix gives~\cite{Lee10}
\begin{equation}
\langle \hat{O}_{\chi}^{B}(\alpha_{\chi};\eta) \rangle_{\eta} = \left\{ \begin{array}{ll} \frac{\pi}{2 \eta} W_{\Phi}^{\eta \, B}(\alpha_{\chi}) + \left( 1 - \frac{1}{\eta} \right)  & \mathrm{if} \ \frac{1}{2} < \eta \leq 1, \\ 2 W_{\Phi}^{\eta \, B}(\alpha_{\chi}) - 1 & \mathrm{if} \ 0\le\eta \leq \frac{1}{2}. \end{array} \right.
\end{equation}
Here, $W_{\Phi}^{\eta \, B}(\alpha_{\chi})$ is the Wigner function of the detected state, which is related to the Wigner function of the initial state before losses $\vert \Phi \rangle_{\chi}^{B}$ by the Gaussian convolution
\begin{equation}
\label{eq:Gaussian_detected_convolution}
\begin{aligned}
W_{\Phi}^{\eta \, B}&(X_{\chi},P_{\chi}) =  \frac{2}{\pi (1-\eta)} \int_{-\infty}^{\infty} \int_{-\infty}^{\infty} dX'_{\chi} dP'_{\chi} \\
&\times W_{\Phi}^{B}(X'_{\chi},P'_{\chi}) 
e^{-2 \left[ \frac{(X_{\chi} - \sqrt{\eta} X'_{\chi})^{2}}{1-\eta} + \frac{(P_{\chi} - \sqrt{\eta} P'_{\chi})^{2}}{1-\eta} \right]}.
\end{aligned}
\end{equation}
The measured Wigner function given in Eq.~(\ref{eq:Gaussian_detected_convolution}) corresponds to the $s$-parametrized quasi-probability distribution $W_{\Phi}^{B},(\alpha_{\chi},s)$, of $\vert \Phi \rangle_{\chi}^{B}$ with $s = - \frac{(1-\eta)}{\eta} $ \cite{Cahi69a,Cahi69b}. Exploiting the properties of such distributions, it is straightforward to prove that~\cite{Lee10} 
\begin{equation}
\vert \langle \hat{O}_{\chi}^{B}(\alpha_{\chi};\eta) \rangle_{\eta} \vert \leq 1
\end{equation}
for all values of $\eta$. We can then define the overall measurement performed on the multiphoton state as
\begin{equation}
\hat{O}_{\chi,\chi_{\bot}}^{B}(\alpha_{\chi},\chi;\eta) = \hat{O}^{B}_{\chi}(\alpha_{\chi};\eta) \otimes \hat{\mathbbm{1}}^{B}_{\chi_{\bot}}
\end{equation}
with average values bounded by $\vert \langle \hat{O}_{\chi,\chi_{\bot}}^{B}(\alpha_{\chi},\chi;\eta) \rangle_{\eta} \vert \leq 1$. The two-mode correlation operator for the entanglement witness is then defined as:
\begin{equation}
\hat{\tilde{C}}^{}(\alpha_{\chi}, \chi; \phi; \eta) = \hat{\sigma}^{A}(\phi) \otimes \hat{O}_{\chi,\chi_{\bot}}^{B}(\alpha_{\chi},\chi;\eta)
\end{equation}
where $\hat{\sigma}^{A}(\phi)$ is the Pauli operator for mode $\mathbf{k}_{A}$ along the direction $\phi$ in the Bloch sphere.
Starting from these definitions, we construct the witness operator
\begin{equation}
\label{eq:witness_op}
\begin{aligned}
\hat{\cal W} &= \hat{\tilde{C}}_{\eta}^{}(\alpha'_{\chi},\chi';\phi') + \hat{\tilde{C}}_{\eta}^{}(\alpha'_{\chi},\chi';\phi) + \\
&+ \hat{\tilde{C}}_{\eta}^{}(\alpha_{\chi},\chi;\phi') - \hat{\tilde{C}}_{\eta}^{}(\alpha_{\chi},\chi;\phi).
\end{aligned}
\end{equation}
In order to define the bounds on $\langle\hat{\cal W}\rangle$ satisfied by separable states, we consider a generic MMS-separable state described by the density matrix $\hat{\rho}^{\mathrm{sep}} = \sum_{i} p_{i} \hat{\rho}^{A}_{i} \otimes \hat{\rho}^{B}_{i}$. After detection losses on the multiphoton mode $\mathbf{k}_{B}$, such state evolves into $\hat{\rho}^{\mathrm{sep}} = \sum_{i} p_{i} \hat{\rho}^{A}_{i} \otimes \hat{\rho}^{\eta \, B}_{i}$, which gives
\begin{equation}
\begin{aligned}
\vert \langle \hat{W} \rangle_{\eta}^{\mathrm{sep}} \vert &= \Big \vert \sum_{i} p_{i} \big ( \langle \hat{A'} \rangle^{i} \langle \hat{B'} \rangle_{\eta}^{i} + \langle \hat{A'} \rangle^{i} \langle \hat{B} \rangle_{\eta}^{i} + \\
&+ \langle \hat{A} \rangle^{i} \langle \hat{B'} \rangle_{\eta}^{i} - \langle \hat{A} \rangle^{i} \langle \hat{B} \rangle_{\eta}^{i} \big ) \Big \vert
\end{aligned}
\end{equation}
where
\begin{equation}
\begin{aligned}
\langle \hat{B} \rangle^{i}_{\eta} &{=}\mathrm{Tr} \left[ \hat{O}_{\chi,\chi_{\bot}}^{B}(\alpha_{\chi},\chi;\eta) \hat{\rho}^{\eta \, B}_{i} \right],\\ 
\langle \hat{A} \rangle^{i} &{=}\mathrm{Tr} \left[ \hat{\sigma}^{A}(\phi) \hat{\rho}^{A}_{i} \right],\\ 
\langle \hat{A'} \rangle^{i}&{=}\mathrm{Tr} \left[ \hat{\sigma}^{A}(\phi') \hat{\rho}^{A}_{i} \right],\\  
\langle \hat{B'} \rangle^{i}_{\eta}&{=}\mathrm{Tr} \left[ \hat{O}_{\chi',\chi'_{\bot}}^{B}(\alpha'_{\chi},\chi';\eta) \hat{\rho}^{\eta \, B}_{i} \right].
\end{aligned}
\end{equation}
As all these terms satisfy $\vert \langle \hat{X} \rangle^{i} \vert \leq 1$ with $\hat{X} = \{ \hat{A}, \hat{A}', \hat{B}, \hat{B}' \} $, we get
\begin{equation}
\vert \mathcal{W}_{\eta}^{sep} \vert \leq 2,
\end{equation}
which is the desired witness condition. We conclude by discussing the features of this inequality. On one side, we note that the derivation of this bound is performed under the assumption that the state is measured with efficiency $\eta$. Hence, such a witness operator permits us to demonstrate the entanglement \textit{before} detection losses. On the other side, no assumption is necessary on the MMS source due to the generality of the derived criterion. Finally, we note that for the case $\eta=1$ this entanglement witness coincides with the CHSH-based inequality of Eq.~(\ref{eq:CHSH_def_2}), given that no assumption is made on the efficiency of the detection apparatus.

\section{Correlator for the CHSH-based test in ideal conditions}

\label{sec:appC}

In this section we report the full calculation of the correlator $\mathcal{C}^{}(X_{\chi},P_{\chi},\chi;\phi)$ reported in the main letter.  We begin with the two-mode correlation $\hat{Q}^{}$, defined as
\begin{equation}
\hat{Q}^{}(\alpha_{\chi},\alpha_{\chi_{\bot}}, \chi;\phi) = \hat{\sigma}^{A}(\phi) \otimes \left( \hat{\Pi}^{B}_{\chi}(\alpha_{\chi}) \otimes \hat{\Pi}^{B}_{\chi_{\bot}}(\alpha_{\chi_{\bot}}) \right)
\end{equation}
This operator corresponds to the measurement of the generalized parity operator on both polarization modes $\{ \vec{\pi}_{\chi}, \vec{\pi}_{\chi_{\bot}} \}$ of the macropart of our state.
The average $\mathcal{Q}^{}(\alpha_{\chi},\alpha_{\chi_{\bot}},\chi;\phi) = \,_{AB}\langle \Psi^{-} \vert \hat{Q}^{} \vert \Psi^{-} \rangle_{AB}$ is related to the correlator of the CHSH-based inequality by
\begin{equation}
\label{eq:integr_C_D}
\mathcal{C}^{}(\alpha_{\chi},\chi;\phi) = \frac{2}{\pi} \int d^{2}\alpha_{\chi_{\bot}} \mathcal{Q}^{}(\alpha_{\chi},\alpha_{\chi_{\bot}},\chi;\phi).
\end{equation}
This expression holds by considering the closure relation $\frac{2}{\pi} \int d^{2}\alpha_{\chi_{\bot}} \hat{\Pi}_{\chi_{\bot}}(\alpha_{\chi_{\bot}}) \equiv \mathbbm{1}_{\chi_{\bot}}$, which in turn comes from the normalization of the Wigner function.

\subsection{Two-mode correlator}
We now calculate the two-mode correlator $\mathcal{Q}^{}(\alpha_{\chi},\alpha_{\chi_{\bot}},\chi;\phi)$.
Let us recall the expression of the micro-macro state under investigation:
\begin{equation}
\vert \Psi^{-} \rangle_{AB} = \frac{1}{\sqrt{2}} (\vert \phi \rangle_{A} \vert \Phi^{\phi}_{\bot} \rangle_{B} - \vert \phi_{\bot} \rangle_{A} \vert \Phi^{\phi} \rangle_{B})
\end{equation}
where the state has been expressed in a generic equatorial polarization basis $\{ \vec{\pi}_{\phi},\vec{\pi}_{\phi_{\bot}}\}$. The value of $\mathcal{Q}^{}(\alpha_{\chi},\alpha_{\chi_{\bot}},\chi;\phi)$ is obtained by exploiting  the relation between the two-mode $\hat{\Pi}^{B}_{\chi}(\alpha_{\chi}) \otimes \hat{\Pi}^{B}_{\chi_{\bot}}(\alpha_{\chi_{\bot}})$ operator and the two-mode Wigner function $_{B}\langle \Phi \vert\hat{\Pi}^{B}_{\chi}(\alpha_{\chi}) \otimes \hat{\Pi}^{B}_{\chi_{\bot}}(\alpha_{\chi_{\bot}}) \vert \Phi \rangle_{B}{=}\frac{\pi^{2}}{4} W_{\Phi}(\alpha_{\chi},\alpha_{\chi_{\bot}})$. We get
\begin{equation}
\mathcal{Q}^{}(\alpha_{\chi},\alpha_{\chi_{\bot}},\chi;\phi) = \frac{\pi^{2}}{8} \left[ W^{B}_{\phi_{\bot}}(\alpha_{\chi},\alpha_{\chi_{\bot}}) - W^{B}_{\phi}(\alpha_{\chi},\alpha_{\chi_{\bot}}) \right].
\end{equation}
\begin{widetext}
Here, $W^{B}_{\phi_{\bot}}$ and $W^{B}_{\phi}$ stand for the two-mode Wigner functions of amplified $\vert \phi_{\bot} \rangle$ and $\vert \phi \rangle$ single-photon states respectively, evaluated at the rotated phase-space variables $\{ \alpha_{\chi}, \alpha_{\chi_{\bot}} \}$. The correlator $\mathcal{Q}^{AB}(\alpha_{\chi},\alpha_{\chi_{\bot}},\chi;\phi)$ is then derived starting from the expression of the Wigner functions~\cite{Spag09} (where $S = \sinh g$ and $C = \cosh g$)
\begin{equation}
\begin{aligned}
W^{B}_{\phi_{\bot}}(\alpha_{\chi},\alpha_{\chi_{\bot}}) &= \frac{4}{\pi^{2}} \left\{ 4 \left[ \vert \alpha_{\phi_{\bot}} \vert^{2} (1 + 2 S^{2}) + 2 C S \ \mathrm{Re}(\alpha_{\phi_{\bot}}^{2} e^{\imath \phi})\right] - 1 \right\} e^{-2 \left[ ( \vert \alpha_{\phi_{\bot}} \vert^{2} + \vert \alpha_{\phi} \vert^{2} ) (1 + 2 S^{2}) + 2 C S \mathrm{Re}(\alpha_{\phi_{\bot}}^{2} e^{\imath \phi} - \alpha_{\phi}^{2} e^{\imath \phi}) \right] }\\
W^{B}_{\phi}(\alpha_{\chi},\alpha_{\chi_{\bot}}) &= \frac{4}{\pi^{2}} \left\{ 4 \left[ \vert \alpha_{\phi} \vert^{2} (1 + 2 S^{2}) + 2 C S \ \mathrm{Re}(\alpha_{\phi}^{2} e^{\imath \phi} )\right] - 1 \right\} e^{-2 \left[ ( \vert \alpha_{\phi_{\bot}} \vert^{2} + \vert \alpha_{\phi} \vert^{2} ) (1 + 2 S^{2}) + 2 C S \mathrm{Re}(\alpha_{\phi_{\bot}}^{2} e^{\imath \phi} - \alpha_{\phi}^{2} e^{\imath \phi}) \right] }
\end{aligned}
\end{equation}
\end{widetext}
by rotating the polarization of the phase-space variables $\{ \alpha_{\phi}, \alpha_{\phi_{\bot}} \}$ as
\begin{equation}
\begin{split}
\alpha_{\phi} &= e^{\imath(\chi-\phi)/2}[ \alpha_{\chi} \cos(\chi-\phi) - \imath \alpha_{\chi_{\bot}} \sin(\chi-\phi)],\\
\alpha_{\phi_{\bot}} &= e^{\imath(\chi-\phi)/2}[ \alpha_{\chi_{\bot}} \cos(\chi-\phi) - \imath \alpha_{\chi} \sin(\chi-\phi)].
\end{split}
\end{equation}
Finally, we replace the complex phase-space variables with the real quadratures $(X_{\chi},P_{\chi},X_{\chi_{\bot}},P_{\chi_{\bot}})$ and obtain the full expression for $\mathcal{Q}^{}(X_{\chi},P_{\chi},X_{\chi_{\bot}},P_{\chi_{\bot}},\chi;\phi)$. However, this is too lengthy and rather uninformative and will not be reported here. 

\subsection{Single-mode correlator}

We now calculate the single mode correlator $\mathcal{C}^{}(X_{\chi},P_{\chi},\chi;\phi)$. The choice of this measurement operator allows us to capture the nonlocal features of the MMS state generated by amplification of an entangled photon pair. To evaluate this quantity we exploit Eq.~(\ref{eq:integr_C_D}),
\begin{equation}
\begin{aligned}
\mathcal{C}^{}(X_{\chi},P_{\chi},\chi;\phi){=}\frac{2}{\pi} \int\! \int d\Omega\, \mathcal{Q}^{}(X_{\chi},P_{\chi},X_{\chi_{\bot}},P_{\chi_{\bot}},\chi;\phi)
\end{aligned}
\end{equation}
where the integral in $d^{2}\alpha_{\chi_{\bot}}$ has been replaced by the integral in the quadrature variables $ d\Omega{=}dX_{\chi_{\bot}} dP_{\chi_{\bot}}$. After straightforward algebra, we obtain the following expression for the correlator
\begin{equation}
\label{eq:correlator_lossless_2}
\begin{aligned}
\mathcal{C}^{}(X_{\chi},P_{\chi},\chi;\phi) &= \cos[2(\chi-\phi)] e^{-2(e^{-2 g} \overline{X}^{2}_{\chi} + e^{2 g} \overline{P}^{2}_{\chi})} \\
& \times \left[ 1 - 2 (e^{-2 g} \overline{X}^{2}_{\chi} + e^{2 g} \overline{P}^{2}_{\chi})\right]
\end{aligned}
\end{equation}
where $\{ \overline{X}_{\chi}, \overline{P}_{\chi} \}$ define a set of rotated variables $\overline{X}_{\chi}= X_{\chi} \cos(\chi/2)-P_{\chi} \sin(\chi/2)$, $\overline{P}_{\chi}=X_{\chi} \sin(\chi/2)+P_{\chi} \cos(\chi/2)$. The maximum of such a correlation operator is obtained at the origin of the phase-space and reads
$\mathcal{C}^{}(0,0,\chi;\phi) = \cos[2(\chi-\phi)]$.

\section{Correlator for the CHSH-based test under detection losses and nonunitary injection efficiency}
\label{sec:appD}

Here we report in details the calculation of the correlator $\mathcal{C}^{}_{p,\eta}$, when detection losses and a nonunitary 
injection efficiency are taken into account. These two effects represent the two main issues for an experimental observation of
entanglement in a MMS.

The model for the effect of losses at the detection stage is performed by inserting a beam-splitter of transmittivity $\eta$ along the transmission path of the field on mode $\mathbf{k}_{B}$. The other port of this beam-splitter is injected with a vacuum state,
thus introducing vacuum-noise fluctuations in the system. Here we demonstrate that the correlator $\mathcal{C}_{\eta}^{}$ in presence 
of detection losses $\eta$ can be evaluated as the convolution of the lossless correlator $\mathcal{C}^{}$ with a Gaussian function 
of the form:
\begin{equation}
\label{eq:convolution_correlator}
\begin{aligned}
\mathcal{C}^{}_{\eta}(X_{\chi},P_{\chi},\chi;\phi) =& \frac{2}{\pi (1-\eta)} \int\int dX'_{\chi} dP'_{\chi} \ C^{}(X'_{\chi},P'_{\chi},\chi; \phi) \\
&\times
e^{-2 \left[ \frac{(X_{\chi} - \sqrt{\eta} X'_{\chi})^{2}}{1-\eta} + \frac{(P_{\chi} - \sqrt{\eta} P'_{\chi})^{2}}{1-\eta} \right]}.
\end{aligned}
\end{equation}
We begin by writing the density matrix $\hat{\rho}_{\eta}^{\Psi^{-}}$ of the micro-macro state after losses occur at the detection stage
\begin{equation}
\begin{aligned}
\hat{\rho}_{\eta}^{\Psi^{-}} {=}& \frac{1}{2} \Big\{ \vert \phi \rangle_{A} \langle \phi \vert{ \otimes }\mathcal{L}\big[ 
\vert \Phi^{\phi_{\bot}} \rangle_{B} \langle \Phi^{\phi_{\bot}} \vert \big] \\
&+ \vert \phi_{\bot} \rangle_{A} \langle \phi_{\bot} \vert { \otimes } \mathcal{L}\big[ \vert \Phi^{\phi} \rangle_{B} \langle \Phi^{\phi} \vert \big]  \\
&- \vert \phi \rangle_{A} \langle \phi_{\bot} \vert { \otimes } \mathcal{L}\big[ \vert \Phi^{\phi_{\bot}} \rangle_{B} \langle \Phi^{\phi} \vert \big] \\
&{-}\vert \phi_{\bot} \rangle_{A} \langle \phi \vert{ \otimes } \mathcal{L}\big[ \vert \Phi^{\phi} \rangle_{B} \langle \Phi^{\phi_{\bot}} \vert \big] \Big\},
\end{aligned}
\end{equation}
where $\mathcal{L}[\cdot]$ is the map that describes the action of detection losses. The evaluation of the correlation operator $\mathcal{Q}^{}$ on this density matrix leads to
\begin{equation}
\label{eq:DABeta_gen}
\mathcal{Q}^{}_{\eta}(\alpha_{\chi},\alpha_{\chi_{\bot}},\chi;\phi) {=} \frac{\pi^{2}}{8} \big[ W^{B}_{\eta,\phi_{\bot}}(\alpha_{\chi},\alpha_{\chi_{\bot}}) \\
- W^{B}_{\eta,\phi}(\alpha_{\chi},\alpha_{\chi_{\bot}}) \big]
\end{equation}
where $W^{B}_{\eta,\phi}$ and $W^{B}_{\eta,\phi_{\bot}}$ are the Wigner functions of the macrostates $\vert \Phi^{\phi} \rangle$ and $\vert \Phi^{\phi_{\bot}}\rangle$ after losses. The action of detection losses in the phase-space can be written in the form of a Gaussian convolution~\cite{Leon93}
\begin{equation}
W_{\eta}(X,P) = \int \int dX' dP' \ W(X,P) K_{\eta}(X,P,X',P'),
\end{equation}
where $K_{\eta}(X,P,X',P'){=}\frac{2}{\pi (1-\eta)} \exp\{-2[\frac{(X-\sqrt{\eta} X')^{2}}{1-\eta}+\frac{(P - \sqrt{\eta} P')^2}{1-\eta}]\}$.
The correlator $\mathcal{C}^{}_{\eta}$ is obtained from $\mathcal{Q}^{}_{\eta}$ as
\begin{equation}
\begin{aligned}
\mathcal{C}^{}_{\eta}(X_{\chi},P_{\chi},\chi;\phi) &= \frac{2}{\pi} \int \int d\Omega\mathcal{Q}^{}_{\eta}(X_{\chi},P_{\chi},X_{\chi_{\bot}},P_{\chi_{\bot}},\chi;\phi).
\end{aligned}
\end{equation}
By writing explicitly the Wigner function after losses as a Gaussian convolution we obtain
\begin{equation}
\label{eq:dim_conv_interm}
\begin{aligned}
\mathcal{C}^{}_{\eta}(X_{\chi},P_{\chi},\chi;\phi) &= \frac{2}{\pi} \int\int dX_{\chi_{\bot}} dP_{\chi_{\bot}} \mathcal{I}(X'_{\chi},P'_{\chi}) 
\end{aligned}
\end{equation}
where 
\begin{equation}
\begin{aligned}
&\mathcal{I}(X'_{\chi},P'_{\chi})=\int\int dX'_{\chi_{\bot}} dP'_{\chi_{\bot}} \mathcal{Q}^{}(X'_{\chi},P'_{\chi},X'_{\chi_{\bot}},P'_{\chi_{\bot}},\chi;\phi) \\
& \times\int \int dX_{\chi_{\bot}} dP_{\chi_{\bot}} K_{\eta}(X_{\chi_{\bot}},P_{\chi_{\bot}},X'_{\chi_{\bot}},P'_{\chi_{\bot}}).
\end{aligned}
\end{equation}
By changing the integration variables as $X_{\chi_{\bot}} \rightarrow \tilde{X}_{\chi_{\bot}} = \frac{X_{\chi_{\bot}} - \sqrt{\eta} X'_{\chi_{\bot}}}{\sqrt{1-\eta}}$, $P_{\chi_{\bot}} \rightarrow \tilde{P}_{\chi_{\bot}} = \frac{P_{\chi_{\bot}} - \sqrt{\eta} P'_{\chi_{\bot}}}{\sqrt{1-\eta}}$, we have the explicit function
\begin{equation}
\begin{aligned}
\mathcal{I}(X'_{\chi},P'_{\chi})& = \frac{2\vert J \vert}{\pi(1-\eta)}\int \int d\tilde{X}_{\chi_{\bot}} d\tilde{P}_{\chi_{\bot}} e^{-2 (\tilde{X}_{\chi_{\bot}}^{2} + \tilde{P}_{\chi_{\bot}}^{2})}\\
&\times \int\int dX'_{\chi_{\bot}} dP'_{\chi_{\bot}} \mathcal{Q}^{}(X'_{\chi},P'_{\chi},X'_{\chi_{\bot}},P'_{\chi_{\bot}},\chi;\phi) ,
\end{aligned}
\end{equation}
where $\vert J \vert = 1-\eta$. Eq.~(\ref{eq:convolution_correlator}) is found by integrating over $d\tilde{X}_{\chi_{\bot}} d\tilde{P}_{\chi_{\bot}}$, 
using Eq.~(\ref{eq:integr_C_D}) to have $\mathcal{I}(X'_{\chi},P'_{\chi}) = \mathcal{C}^{}(X'_{\chi},P'_{\chi},\chi;\phi)$
and replacing this in Eq.~(\ref{eq:dim_conv_interm}).

We now proceed with the explicit calculation of Eq.~(\ref{eq:convolution_correlator}). As a first step, we rotate the quadratures $(X_{\chi},P_{\chi})$ and the integration variables $(X'_{\chi},P'_{\chi})$ as
\begin{equation}
\begin{aligned}
\overline{\cal X}_{\chi} &{=} {\cal X}_{\chi} \cos(\chi/2) {-} {\cal P}_{\chi} \sin(\chi/2),\\
\overline{\cal P}_{\chi} &{=} {\cal X}_{\chi} \sin(\chi/2) {+} {\cal P}_{\chi} \cos(\chi/2) 
\end{aligned}
\end{equation}
with ${\cal X}=(X,X')$ and ${\cal P}=(P,P')$ and the convention that only primed (unprimed) variables are involved in the equations above.  The correlator $\mathcal{C}^{}_{\eta}$ can be then expressed as a function of the rotated variables. After replacing the expression of $K_{\eta}$ in the correlator $\mathcal{C}^{}_{\eta}$, it is matter of some straightforward (although tedious) algebra to find that 
\begin{equation}
\begin{aligned}
&\mathcal{C}^{}_{\eta}(\overline{X}_{\chi},\overline{P}_{\chi},\chi;\phi) = \frac{\cos[2(\chi-\phi)]e^{-2 \left[ \frac{\overline{X}_{\chi}^{2}}{{\cal M}} + \frac{\overline{P}_{\chi}^{2}}{{\cal N}} \right]}  }{\sqrt{1+4\eta (1-\eta)\overline{n}}}\\
& \left\{ 1 - \frac{(1-\eta) (1 + 2 \eta \overline{n})}{1+4\eta(1-\eta)\overline{n}}- 2 \eta \left[ \frac{e^{2 g} \overline{X}_{\chi}^{2}}{{\cal M}^{2}} + \frac{e^{-2 g} \overline{P}_{\chi}^{2}}{{\cal N}^{2}} \right] \right\}
\end{aligned}
\end{equation}
with ${\cal M}=\eta e^{2 g} + (1-\eta)$ and ${\cal N}=\eta e^{-2 g} + (1-\eta)$.
This expression is maximized at the origin of the phase space, reading
\begin{equation}
\mathcal{C}^{}_{\eta}(\overline{X}_{\chi},\overline{P}_{\chi},\chi;\phi) = \cos[2(\chi-\phi)] \mathcal{L}(\eta,g)
\end{equation}
where the loss function $\mathcal{L}(\eta,g)$ has the form:
\begin{equation}
\mathcal{L}(\eta,g) = \frac{\eta + 2 \eta (1-\eta) \overline{n}}{(1+ 4 \eta (1-\eta) \overline{n})^{3/2}}.
\end{equation}

In typical experimental conditions, the injection of the single photon of the entangled pair $\vert \psi^{-} \rangle_{AB}$ into the OPA occurs with an efficiency $p<1$ because of the imperfect matching between
the optical modes of the amplifier and the single-photon one. Such nonideality can be modeled by allowing for a probability $p$ of correct single-photon injection and a complementary probability $(1-p)$ that just vacuum state is injected in the amplifier and no correlations
between the two output modes are set. This modifies the density matrix of the output modes as
\begin{equation}
\hat{\rho}_{p}^{\psi^{-}} = p \vert \psi^{-} \rangle_{AB} \langle \psi^{-} \vert + (1-p) \frac{\hat{\mathbbm{1}}_{A}}{2} \otimes \vert 0 \rangle_{B} \langle 0 \vert,
\end{equation} 
where $\hat{\mathbbm{1}}_{A} = \vert H \rangle_{A} \langle H \vert + \vert V \rangle_{A} \langle V \vert$ is a completely mixed single-photon polarization
state, and $\vert 0 \rangle_{B} \langle 0 \vert$ is the vacuum state. 
The bipartite state after the amplification process then reads
\begin{equation}
\hat{\rho}_{p}^{\Psi^{-}} = p \vert \Psi^{-} \rangle_{AB} \langle \Psi^{-} \vert + (1-p) \frac{\hat{\mathbbm{1}}_{A}}{2} \otimes \left( \hat{U}_{OPA} \vert
0 \rangle_{B} \langle 0 \vert \hat{U}_{OPA}^{\dag} \right).
\end{equation}
We can now proceed with the calculation of ${\cal C}(\alpha_{\chi},\chi;\phi) $ as
\begin{equation}
\begin{aligned}
& {\cal C}^{}(\alpha_{\chi},\chi;\phi) = p \, _{AB}\langle \Psi^{-} \vert\hat{\sigma}^{A}(\phi) \otimes \hat{\Pi}^{B}(\alpha_{\chi},\chi) 
\vert \Psi^{-} \rangle_{AB} \\
&{+} (1{-}p) \mathrm{Tr}\!\left[ \frac{\hat{\mathbbm{1}}_{A}}{2}{\otimes}\left( \hat{U}_{OPA} \vert
0 \rangle_{B} \langle 0 \vert \hat{U}_{OPA}^{\dag} \right)\hat{\sigma}^{A}(\phi){\otimes}\hat{\Pi}^{B}(\alpha_{\chi},\chi)\right].
\end{aligned}
\end{equation}
As the second term factorizes (due to the lack of quantum correlations) and $\mathrm{Tr} \left[ \frac{\hat{\mathbbm{1}}_{A}}{2} \hat{\sigma}^{A}(\phi) \right] =0$, such contribution is null.
Therefore, under nonideal injection efficiency, the correlator is related to the ideal one according to $\mathcal{C}^{}_{p}(X_{\chi},P_{\chi},\chi;\phi) =p \ \mathcal{C}^{}(X_{\chi},P_{\chi},\chi;\phi)$. This result 
can be extended to the case of nonunitary detection efficiency, leading to
\begin{equation}
\mathcal{C}^{}_{\eta,p}(X_{\chi},P_{\chi},\chi;\phi) = p \ \mathcal{C}^{}_{\eta}(X_{\chi},P_{\chi},\chi;\phi).
\end{equation}

\section{Correlator for the entanglement witness after detection losses and nonunitary injection efficiency}

\label{sec:appE}

Here we sketch the steps needed for the calculation of the correlator $\tilde{\mathcal{C}}^{}_{p,\eta}$ entering the entanglement test based on the witness operator of Eq.~(\ref{eq:witness_op}) under losses and nonideal photon injection. By using arguments similar to those put forward in the previous sections, we have 
\begin{equation}
\begin{aligned}
{\tilde{\cal C}}^{}_\eta(\alpha_{\chi},\chi;\phi)&= \frac{1}{2}\left\{ \mathrm{Tr} \left[ \mathcal{L}\big[ 
\vert \Phi^{\phi_{\bot}} \rangle_{B} \langle \Phi^{\phi_{\bot}} \vert \big] \hat{O}_{\chi,\chi_{\bot}}^{B}(\alpha_{\chi},\chi;\eta) \right] \right. \\ &- \left.  \mathrm{Tr} \left[ \mathcal{L}\big[ 
\vert \Phi^{\phi} \rangle_{B} \langle \Phi^{\phi} \vert \big] \hat{O}_{\chi,\chi_{\bot}}^{B}(\alpha_{\chi},\chi;\eta) \right] \right\},
\end{aligned}
\end{equation}
where ${\cal L}[\cdot]$ is the map describing the lossy process. We focus on the case $\eta \geq \frac{1}{2}$. By exploiting results that have been previously obtained here, we have
$\tilde{\mathcal{C}}^{}_{\eta}(\alpha_{\chi},\chi;\phi){=}\frac{\pi}{4\eta} \int\!d^{2}\alpha_{\chi_{\bot}} \left( W_{\phi_{\bot}}^{\eta}(\alpha_{\chi},\alpha_{\chi_{\bot}}){-}W_{\phi}^{\eta}(\alpha_{\chi},\alpha_{\chi_{\bot}}) \right)$.
We now exploit the chain of relations 
\begin{equation}
\begin{aligned}
&\frac{\pi}{4} \int d^{2}\alpha_{\chi_{\bot}} \left( W_{\phi_{\bot}}^{\eta}(\alpha_{\chi},\alpha_{\chi_{\bot}}) 
-  W_{\phi}^{\eta}(\alpha_{\chi},\alpha_{\chi_{\bot}}) \right) \\
&= \frac{2}{\pi} \int d^{2}\alpha_{\chi_{\bot}} \frac{\pi^{2}}{8} \left( W_{\phi_{\bot}}^{\eta}(\alpha_{\chi},\alpha_{\chi_{\bot}}) 
-  W_{\phi}^{\eta}(\alpha_{\chi},\alpha_{\chi_{\bot}}) \right) &\\
&= \frac{2}{\pi} \int d^{2}\alpha_{\chi_{\bot}} \mathcal{Q}^{}_{\eta}(\alpha_{\chi},\alpha_{\chi_{\bot}},\chi;\phi) = \mathcal{C}^{}_{\eta}(\alpha_{\chi},\chi;\phi)
\end{aligned}
\end{equation}
so as to get
$\tilde{\mathcal{C}}^{}_{\eta}(\alpha_{\chi},\chi;\phi;\eta) = \frac{1}{\eta} \mathcal{C}^{AB}_{\eta}(\alpha_{\chi},\chi;\phi)$.
With an analogous procedure, we obtain
\begin{equation}
\label{eq:correlator_witness_losses}
\tilde{\mathcal{C}}^{}_{\eta}(\alpha_{\chi},\chi;\phi;\eta) = \left\{ \begin{array}{ll} \frac{1}{\eta} \ \mathcal{C}^{}_{\eta}(\alpha_{\chi},\chi;\phi) & \mathrm{if} \ \frac{1}{2} < \eta \leq 1, \\ 2 \ \mathcal{C}^{}_{\eta}(\alpha_{\chi},\chi;\phi) & \mathrm{if} \ \eta \leq \frac{1}{2}. \end{array} \right.
\end{equation}
We can further generalize this result so as to take into account the effect of a nonunitary injection efficiency and finally get 
$\tilde{\mathcal{C}}^{}_{\eta,p}(\alpha_{\chi},\chi;\phi;\eta) = p \ \tilde{\mathcal{C}}^{}_{\eta}(\alpha_{\chi},\chi;\phi;\eta)$.

%\bibliography{bibliography_non_loc_CV}
%\bibliographystyle{apsrev4-1}

%

\end{document}